\documentclass[preprint,superscriptaddress,aps]{revtex4}
\usepackage{amsfonts,amsmath,amssymb}
\usepackage{amsfonts}
\usepackage{graphics}
\usepackage{graphicx}
\usepackage{epsf}

\newcommand{\bea}{\begin{eqnarray}}
\newcommand{\eea}{\end{eqnarray}}

\begin{document}












\title{Correlations between emission events in Rainbow Gravity}

\author{D. A. Gomes\footnote{E-mail: deboragomes@fisica.ufc.br}}
\affiliation{Universidade Federal do Cear\'{a} (UFC), Departamento do F\'{i}­sica - Campus do Pici, Fortaleza, CE, C. P. 6030, 60455-760, Brazil.}

\author{F. C. E. Lima\footnote{E-mail: cleiton.estevao@fisica.ufc.br}}
\affiliation{Universidade Federal do Cear\'{a} (UFC), Departamento do F\'{i}­sica - Campus do Pici, Fortaleza, CE, C. P. 6030, 60455-760, Brazil.}

\author{C. A. S. Almeida\footnote{E-mail: carlos@fisica.ufc.br}}
\affiliation{Universidade Federal do Cear\'{a} (UFC), Departamento do F\'{i}­sica - Campus do Pici, Fortaleza, CE, C. P. 6030, 60455-760, Brazil.}

\begin{abstract}
In this work, we study emission correlations in Rainbow Gravity (RG) black holes known to have black hole remnants. Non-thermal corrections are responsible for generating non-zero correlations between particle emission events during the black hole evaporation process. With this in mind, we calculate the temperatures considering back-reaction effects. Then, we obtain the correlations between emission events for a RG metric proposed by Magueijo and Smolin. We also discuss through a numerical analysis the emission correlations behavior in the last stages of evaporation of the black hole.
\end{abstract}
\maketitle



\section{Introduction}

Due to observational evidence, \cite{akiyama} and the efficiency of theoretical results, new researchers and admirers have become adept at studying astrophysical objects such as black holes. These objects are structures predicted by Einstein's theory of relativity. Initially, many researchers believed that black holes were structures that only absorbed matter with the arising of this theory. However, around the 1970s, the well-known physicist Steven Hawking changed the paradigms of the theory and demonstrated that black holes emit radiation, which was named after him. Hawking's theory has shown that black holes emit radiation at temperatures proportional to their surface gravity due to quantum effects. This seemingly strange result opened new horizons for the study of these amazing and interesting structures.

Since the 1970's, many scholars have tried to understand and accurately describe such objects. We can easily see this from the works in the literature. For instance, in 1973, Bardeen, Carten, and Hawking \cite{2} attempted to discuss in the paper {\it The Four Laws of Black Hole Mechanics} the laws governing black hole dynamics and their analogies with the laws of thermodynamics. Meanwhile, as early as 1973, Bekenstein sought to understand and interpret the study of black hole entropy. Subsequently, Bekenstein himself seeks a generalization to the second law of thermodynamics in order to give an accurate description of such structures. Building on many of these theories that emerged in the mid-1970s, new studies and new lines of research eventually emerged over time. These studies include: information loss and entropy conservation in quantum corrected Hawking radiation \cite{Chen}; the study of Hawking radiation with logarithmic correction tunneling \cite{medved}; the study of quasinormal frequencies of self-dual black holes \cite{vitor}, etc.

One new area of research in black hole theory involves the violation of the Lorentz Symmetry. Such symmetry is fundamental for two of the most successful theories in the last century: General Relativity and Standard Model. It has been one of the most tested symmetries of Nature, and to the better of our knowledge there is no experimental evidence disproving its validity \cite{exp,datatables}. In spite of this, Lorentz symmetry breaking could reveal interesting aspects of nature. For instance, it is believed that a fundamental theory involving General Relativity and Quantum Mechanics would hold at the Planck scale. However, it is not possible to test this hypothesis with current technology. One way to address this problem is to identify Planck suppressed signals of this underlying theory at low energy scales, via Lorentz Symmetry Violation \cite{kostelecky,bluhm}.

One of the approaches to violate the Lorentz invariance is to modify the standard dispersion relation $E^2 - p^2 = m^2$, which is valid in the ultraviolet limit \cite{hooft,amelino,iengo}. This leads to the so-called modified dispersion relations (MDR), which are often associated with the existence of a maximum energy scale. In this context, the special relativity can be extended to include MDR, leading to the Doubly Special Relativity (DSR) \cite{dsr1,dsr2,dsr3}. The term doubly comes from the fact that DSR has two universal constants: the speed of light $c$ and the Planck energy $E_P$.

The extension for DSR in curved spaces, proposed by Magueijo and Smolin, is called Doubly General Relativity or, more popularly, Rainbow Gravity \cite{Magueijo}. In Rainbow Gravity (RG), the spacetime is energy-dependent, i.e., particles with different energies would perceive the spacetime background differently. In this sense, we have a ``rainbow" of metrics defined by one parameter, the ratio between the energy of the test particle and the Planck energy ($E/E_P$). In addition to this, the MDR in Rainbow Gravity are modified by correction terms, known as rainbow functions, that depend on $E/E_P$. 

Many interesting applications of Rainbow Gravity can be found in a variety of contexts as inflation \cite{inflation,inflation2}, branes \cite{branes}, wormholes \cite{wormholes}, avoidance of the big bang singularity \cite{bb,bb2,bb3}, explanation for the absence of black hole detection at LHC \cite{BHatLHC}, among others. However, one of the most fruitful contexts in which Rainbow Gravity has been applied is black hole thermodynamics \cite{bhrg,bhrg1,bhrg2,bhrg3,bhrg4,bhrg5,bhrg6,rainbowthermo}, specially due the existence of the so called black hole remnants \cite{Remnants,Remnants2,Remnants3}.

The black hole remnants are particularly interesting because they represent a possible solution for the information paradox. With this in mind, we will work with the rainbow functions proposed by Magueijo e Smolin \cite{dsr3}, which are known to produce black hole remnants \cite{rainbowthermo}. This work aims to investigate the correlation between two successive particle emission events considering back-reaction effects.

In principle, a black hole with mass, angular momentum, and charge has many unobservable internal configurations that reflect the possible initial configurations of the matter that collapsed to produce the black hole \cite{Hawking}. One parameter for studying this astrophysical object is the entropy of the black hole, which gives us a measure of information about the initial state that was lost in the formation of the black hole \cite{Hawking, Hawking1}. In the hypothesis that entropy is finite, it can be deduced that black holes must emit thermal radiation at some temperature different from zero, that is, Hawking radiation \cite{Hawking}. In this scenario, the black hole emits thermal radiation. The correlations of events have the role of describing the entanglement between the particles leaving Hawking's radiation and its falling partners \cite{Unruh}. In this way, we can understand correlation as the measurement of how entangled these particles are.

The fact is that in recent years there has been a significant increase in discussions of rainbow gravity and studies of black hole thermodynamics, e. g., study of black holes in acceleration in anti-de Sitter \cite{Anabalon} spacetime, investigation of Hawking radiation and deflection of light in modified Schwarzschild black holes \cite{Sakalli}, study of particles tunnelling into black hole \cite{Sakalli1, Sakalli2, Ovgun}, study of the quantum gravitational effect on Hawking radiation from rotating acoustic black holes \cite{Sakalli3}, etc.

In the search for a quantum theory of gravity, it is believed that Planck's energy has a threshold role that separates the classic description from the quantum description \cite{smolin}. In a scenario that the energy of the model is of the order of Planck's energy, we hope that a radically new image of spacetime is necessary. Several candidates for this description are being studied including loop quantum gravity \cite{rovelli,carlip}, string theory \cite{for}, non-commutative geometry \cite{connes}, etc. We motivate our study by the fact that recent astronomical and cosmological observations make it possible to probe effects at the limit where Planck's length is equal to the inverse of Planck's energy, i. e., around the classical limit. These possibilities include possible changes in the energy-momentum relationship, e. g., the dispersion relation used by Mangueijo and Smolin. For the choice of this model, we assume functions for which we have the standard energy-momentum dispersion ratio in a low energy scale. We must emphasize that the rainbow function is not unique, and it is possible to find a series of expressions for the rainbow functions depending on different phenomenological motivations.

In this work, we will investigate for the first time the influence of Rainbow gravity on the particle emission correlation of a black hole. This paper is organized into four sections, starting by discussing the thermodynamic properties of a black hole in a rainbow gravity scenario. In the next section, we investigate the study of correlations between particle emission events. To conclude, we make a numerical analysis of the theoretical results presented in the previous sections and present the physical results found.


\section{The thermodynamic properties for the black hole in rainbow gravity}

Motivated by the work of Magueijo and Smolin \cite{Magueijo} we used the dispersion relation given by
\begin{eqnarray}
\frac{E^{2}}{(1-\gamma E/E_{p})^{2}}-p^{2}=m^{2},
\end{eqnarray}
which leads to the following metric:
\begin{equation}
ds^2 = \left(1 - \gamma \frac{E}{E_p} \right)^2 \left(1 - \frac{2M}{r} \right) dt^2 - \left(1 - \frac{2M}{r} \right)^{-1}dr^{2} - r^2(d \theta^2 + \sin ^2 \theta d\phi ^2).
\end{equation}

Now, we will obtain the temperature and entropy for the metric above. For this purpose, we will make use of the tunneling method based on the works \cite{a,a1,a2,a3,a4}. For a non-rotating black hole, we can write the metric as follows

\begin{equation}
    ds^2 = -f(r)dt^2 + g(r) dr^2 + h(r)(d\theta^2 + \sin^2  \theta d \phi ^2).
\end{equation}
The temperature for the black hole can be easily found via the expression 
\begin{equation}
\label{T1}
    T = \frac{\sqrt{f'(r_+)g'(r_+)}}{4 \pi},
\end{equation}
where $r_+$ is the event horizon radius, while the entropy is obtained from the relation $TdS = dM$. The equation \eqref{T1} can be obtained, via tunnelling method, by comparing the tunnelling probability $\Gamma$ with $e^{-\beta \omega}$, where $\beta = T^{-1}$ ($k_B = 1$).

There are several papers in the literature that use different forms of rainbow functions, e. g., in some investigations of thermodynamics of black holes \cite{rainbowthermo,L2},  or in the study of the uncertainty principle in non-local gravity \cite{Refy}. In our work, we assume that the rainbow functions are 

\begin{eqnarray}
f(r) &=& -\left(1 - \frac{2M}{r} \right)\left(1 - \gamma \frac{E}{E_p} \right)^2; \nonumber \\
g(r) &=& -\left(1 - \frac{2M}{r} \right).
\end{eqnarray}

Since the event horizon is the same as Schwarzschild's one, $r_{\scriptscriptstyle H} = 2M$, we arrived at the result:

\begin{equation}
\label{tempe}
T =  \left(1 - \gamma \frac{E}{E_p} \right) T_0,
\end{equation}
where $T_0  = (8 \pi M)^{-1}$ is the Schwarzschild temperature.

Since Hawking radiation emission is a quantum process, the quanta emitted must obey the Heisenberg uncertainty principle, which must be valid in GR \cite{HUP, HUP2}, so that $\Delta x\Delta p \geqslant 1$, in natural units. We can obtain from the uncertainty principle a minimum energy value $E\geqslant 1 /\Delta x$, where $E$ is the particle energy emitted in the Hawking radiation process. Near the event horizon we have $\Delta x \approx r_{\scriptscriptstyle H}=2M$ \cite {Remnants2}. Then,

\begin{equation}
E \geqslant 1/2M.
\end{equation}

Keeping in mind that in natural units we have $ E_p = 1 $, we can rewrite the surface gravity and black hole temperature as \cite{rainbowthermo}
\begin{equation}\label{eq:36}
T =\left(1 -  \frac{\gamma}{2M} \right) T_0.
\end{equation}

From equation \eqref{eq:36}, it is easy to see that the \textit{rainbow} $\gamma$ parameter is responsible for modifying temperature of the Schwarzschild black hole and, consequently, its surface gravity. In addition, the usual results are recovered at the limit $\gamma\rightarrow 0$. Note that since $\gamma>0$, the obtained temperature is lower than in the usual Schwarzschild case, as we will see numerically in later sections.


The entropy for this model is given by \cite{rainbowthermo}
\begin{eqnarray}
\label{entropy_n}
S &=& 16 \pi \int \frac{M^2}{2M - \gamma}dM 
\nonumber \\
&=& S_0 + 2 \pi \gamma [2M + \gamma \ln(2M - \gamma )],
\end{eqnarray}
where $S_0  = 4 \pi M^2$ is the Schwarzschild entropy.

We can notice that $S>S_0$, in agreement with the decreasing in temperature. Furthermore, we see that at the limit $\gamma\rightarrow 0$, the Schwarzschild entropy $S_0$ is recovered, especially for large values of $M$, as shown in the numerical result in section 4. We will also see that the entropy $S$ becomes much larger than $S_0$ when $M$ is approximately of the order of $\gamma$ and that for large values of $M$, $S$ does not differ much from $S_0$.



\section{Correlation between emission events}

In physics problems, for the sake of simplicity, we often consider a particle to have no mass, which can be translated as dealing with a probe particle or neglecting back reaction effects. That is precisely the case of eq. (\ref{T1}). The tunnelling method, which is used in this work, relies on the tunnelling probability of a particle through the black hole event horizon. We can see from eq. (\ref{eq:36}) that the black hole temperature depends on its mass $M$. In other words, we considered the temperature for a fixed spacetime background. However, once a particle is emitted from the black hole, it carries away an energy $\omega$, decreasing the black hole mass to $M-\omega$ due to energy conservation. Consequently, the black hole temperature changes continuously as the black hole radiates. Therefore, for a more realistic analysis of the black hole temperature due to the particle emission, we should calculate the temperature considering the back-reaction effects.

We will now calculate the temperature back-reaction correction and the correlation between emission events for the Schwarzschild metric proposed by Magueijo and Smolin \cite{Magueijo}. The self-gravitational effects in the quantum tunnelling formalism were investigated in Refs \cite{parikh,vagenas}.

We will consider that a particle with energy $\Tilde{\omega}$ is emitted by the black hole, in which case that spacetime background becomes
\begin{equation}
    ds^2 = -f[r(M-\Tilde{\omega})]dt^2 + g[r(M-\Tilde{\omega})] dr^2 + h[r(M-\Tilde{\omega})](d\theta^2 + \sin^2  \theta d \phi ^2).
\end{equation}
It should be noted that $r$ is no longer a function of $M$ but rather a function of $(M-\Tilde{\omega})$, where $M$ is the black hole mass before the emission of the particle. To avoid a violation of the uncertainty principle, we assume that the transition $r(M) \rightarrow r(M-\Tilde{\omega})$ is smooth, while $(\Tilde{\omega})$ goes from $0$ to $\omega$.

With this in mind, we see that the term $\beta \omega$ in $\Gamma = e^{-\beta \omega}$ changes as
\begin{eqnarray}
 \int_0^{\omega} \beta (M-\Tilde{\omega}) d\Tilde{\omega} &\approx& 
    \int_0^{\omega} [\beta(M) - \Tilde{\omega} \partial_M \beta(M) + \mathcal{O}(\Tilde{\omega}^2) ]d \Tilde{\omega} \nonumber \\
  & =& \beta(M) \left[ \omega - \frac{\omega}{2} \frac{\partial_M \beta(M)}{\beta(M)} + \mathcal{O}(\Tilde{\omega}^3)\right],
\end{eqnarray}
where $\partial_M$ denotes the derivation with respect to the mass $M$. From the first black hole thermodynamics, $TdS = dM$, we can see that $\beta = \partial_M S$. In this case, the last equation becomes
\begin{equation}
 \int_0^{\omega} \beta (M-\Tilde{\omega}) d\Tilde{\omega} = \omega \partial_M S - \frac{\omega^2}{2} \partial_M^2 S + \mathcal{O}(\Tilde{\omega}^3)
 \approx -[ S(M-\omega) - S(M)].
 \end{equation}
Now, we can see that the tunnelling probability is then given by $\Gamma = \exp[S(M-\omega) - S(M)] = e^{\Delta S}$ and, consequently, $T = \omega/\Delta S$.

Writing the eq. (\ref{entropy_n}) as a function of mass, we will have that
\begin{eqnarray}
S(M) = 4\pi M^2 + 2 \pi \gamma [2M + \gamma \ln(2M - \gamma )].
\end{eqnarray}
Consequently, the entropy for the black hole after the emission of a particle with energy $\omega$ is given by
\begin{equation}
S(M - \omega) = 4\pi (M - \omega)^2 + 2 \pi \gamma [2(M - \omega) + \gamma \ln(2(M - \omega) - \gamma )].
\end{equation}

Therefore, the variation of the entropy before and after the emission is
\begin{eqnarray}
\Delta S &=& 
4\pi \omega (\omega - 2M) + 2 \pi \gamma \left[-2 \omega + \gamma \ln \left( \frac{2(M - \omega) - \gamma}{2M - \gamma} \right) \right].
\end{eqnarray}

\noindent The corrected temperature will be given by 
\begin{equation}
\label{temperatureco}
T = \left[
4\pi (2M -\omega + \gamma) - 2 \pi \gamma^2 \omega^{-1}\ln \left( \frac{2(M - \omega) - \gamma}{2M - \gamma} \right)
\right]^{-1},
\end{equation}
where we made $k_B=1$.
Note that when the rainbow parameter tends to zero ($\gamma\rightarrow 0$), we recover Schwarzschild's corrected temperature $T=[8\pi (M-\omega/2)]^{-1}$, as expected.

The correlation between emission events is given by
\begin{equation} 
C(\omega_1 + \omega_2; \omega_1 , \omega_2) = \ln \Gamma (M, \omega_1 + \omega_2) - \ln [\Gamma (M, \omega_1)\Gamma (M,\omega_2)].
\end{equation}

For our case, we have 
\begin{eqnarray}
\Gamma (M, \omega_1 + \omega_2) &=& 
\exp \bigg\{ 4\pi [-(2M + \gamma)(\omega_1+ \omega_2) + (\omega_1+ \omega_2)^2 ]  \nonumber \\
&+& \left. 2 \pi \gamma^2 \ln \left[ \frac{2(M - \omega_1- \omega_2) - \gamma}{2M - \gamma} \right] 
\right\rbrace.
\end{eqnarray}
Therefore, 

\begin{equation}
C =
8 \pi \omega_1 \omega_2 
+ 2 \pi \gamma^2 \ln \left\lbrace \frac{[2(M - \omega_1- \omega_2) - \gamma][2M - \gamma]}{[2(M - \omega_1) - \gamma][2(M -  \omega_2) - \gamma]}
\right\rbrace.
\end{equation}

Note that, in the limit $\gamma\rightarrow 0$, we have recovered the Schwarzschild black hole correlation $C_0=8\pi\omega_1\omega_2$. Also, for small values of $\gamma$, the correlations $C$ and $C_0$ become approximately equal as the mass $M$ becomes very large.


\section{Numerical Results and Discussions}

\begin{figure}[h!]
\centering
\includegraphics[scale=0.8]{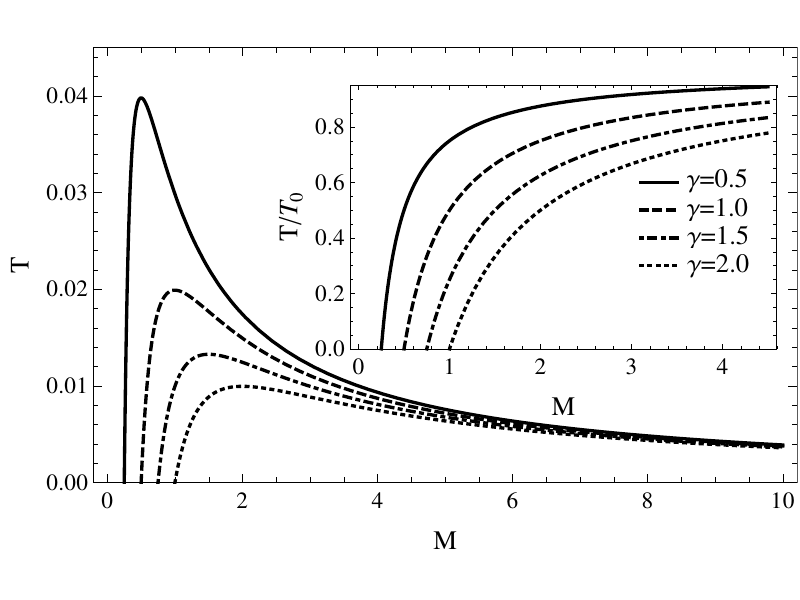}
\vspace{-0.2cm}
\caption{Behavior of temperature and temperature ratio for various values of the rainbow parameter.}
\label{temp}
\end{figure}

\begin{figure}[h!]
\centering
\includegraphics[scale=0.8]{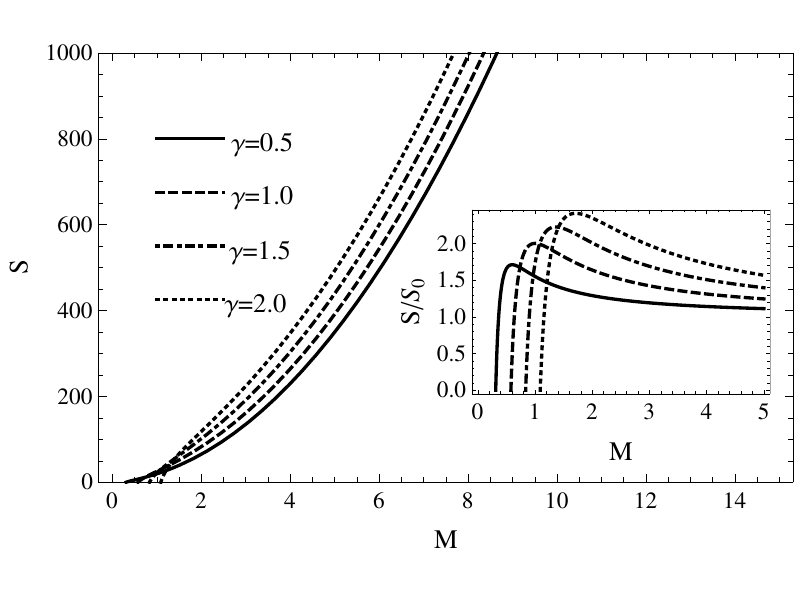}
\vspace{-0.2cm}
\caption{Behavior of entropy and entropy ratio for various values of the rainbow parameter.}
\label{entropy}
\end{figure}

In this section, we will turn our attention to the study of numerical and graphical results for a black hole in the rainbow gravity scenario. In our results, we will consider only values of mass greater than $\gamma/2$, which corresponds to the remnant radius of the black hole. Using the analytical results expressed in eq. (\ref{eq:36}) and in eq. (\ref{entropy_n}), we immediately obtain the results presented in the respective Fig. [\ref{temp}] and Fig. [\ref{entropy}].

From Fig.  [\ref{temp}] we clearly notice that the temperature tends to zero when $M\rightarrow\gamma/2 $. The same behaviour is observed for $T/T_0$. Meanwhile, the temperature value is maximal in $M = \gamma$ and tends to zero as $M \rightarrow \infty$, independently of the rainbow parameter $\gamma$. This is exactly what is expected for a Schwarzschild black hole and happens due to the fact that Schwarzschild's temperature $T_0$ drops faster than $T$. From this, we can clearly see that the rainbow parameter directly influences the behavior of the thermodynamic properties of the model. 

For the entropy, in agreement with the behavior of the temperature, we realized that, as shown in Fig. [\ref{entropy}], the entropy $S$ tends to zero before M becomes $\gamma/2$. Along with this behaviour of the entropy, the heat capacity is also zero for $M = \gamma/2$, as shown in Ref. \cite{rainbowthermo}. This means that the black hole stops radiating for this value of mass, which indicates that the model has a remnant.  Also, as the temperature decreases, the model entropy increases monotonically, as expected. 

\begin{figure}[h!]
\centering
\includegraphics[scale=0.8]{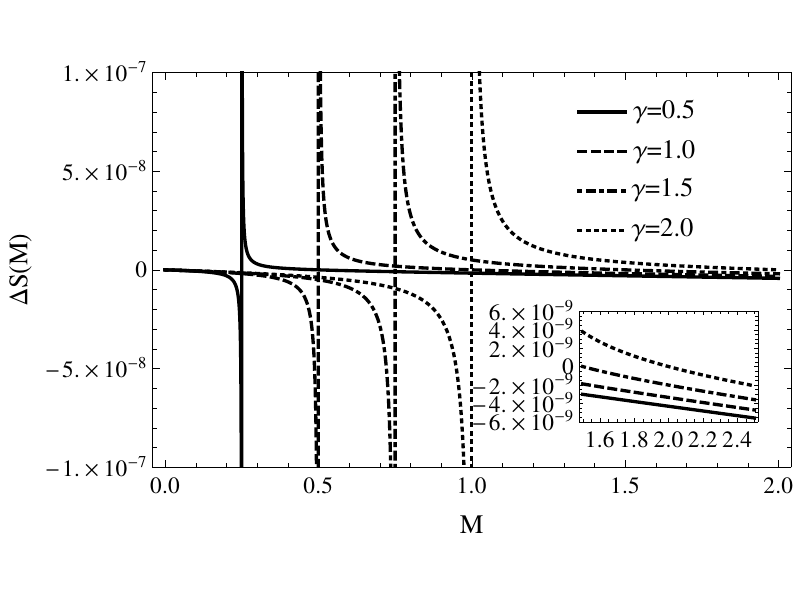}
\vspace{-0.2cm}
\caption{Entropy variation as a function of mass for several values of rainbow parameter and particle with same energy.}
\label{variacaoent1}
\end{figure}

\begin{figure}[h!]
\centering
\includegraphics[scale=0.8]{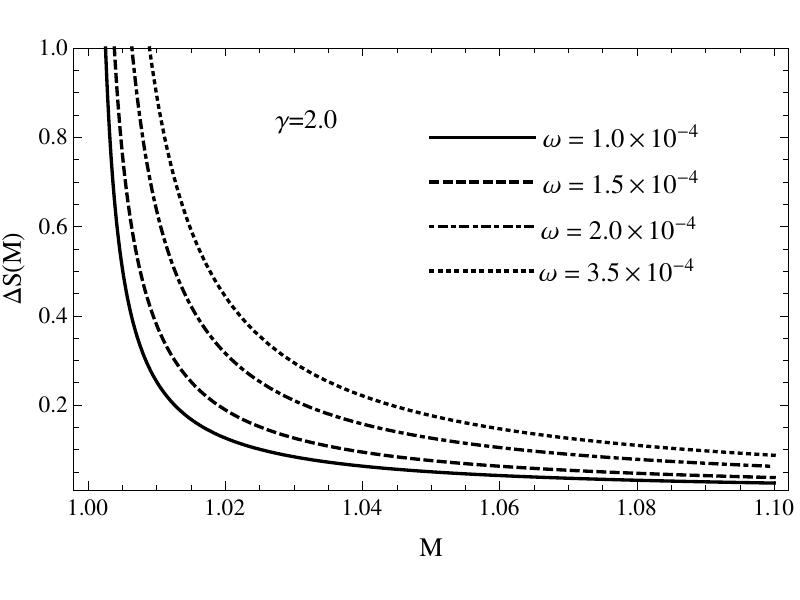}
\vspace{-0.2cm}
\caption{Entropy variation as function of mass for a constant rainbow parameter and particles with different energies.}
\label{variacaoent2}
\end{figure}

We now turn our attention to the behavior of the entropy variation for various values of the rainbow parameter. The results obtained are presented in the Figs. [\ref{variacaoent1}] and [\ref{variacaoent2}]. For a better understanding of our results, the next graphic begin in $M=0$.
Analyzing the entropy variation as a function of the mass for several rainbow parameter values, we notice that when $M>>1$, the entropy correction by the back-reaction effect is negligible, so $\Delta S\rightarrow 0$ when $M\rightarrow +\infty $. As can be observed from Fig. [\ref{variacaoent1}], the back-reaction effect is considerable only for the $\frac{\gamma}{2}<M<\gamma$ region. Also, it is worth to mention the existence of a formation law for entropy variation as a function of the rainbow parameter, given by
\begin{eqnarray}
\label{formation}
\Delta S(M)\varpropto \tan\bigg(\frac{M\pi}{\gamma}\bigg), \, \, \, \, \, \text{para} \, \, \, \, \, 0<M<\gamma,
\end{eqnarray}
for all $\gamma$. 

In order to understand how the energy variation of the emitted particles changes the entropy variation and, consequently, the tunneling probability, we will investigate how $\Delta S$ behaves when $\omega$ varies. Keeping this in mind, we obtain the graphical result shown in Fig. [\ref{variacaoent2}]. From this figure, we can observe that the entropy variation and the tunneling probability become more significant for particles with higher energies. Furthermore, as $M \rightarrow \gamma /2$, the entropy variation increases significantly. This means that the tunneling probability of a particle with energy $\omega$ gets higher and higher as the we approach the remnant state of the black hole.

\begin{figure}[h!]
\centering
\includegraphics[scale=0.8]{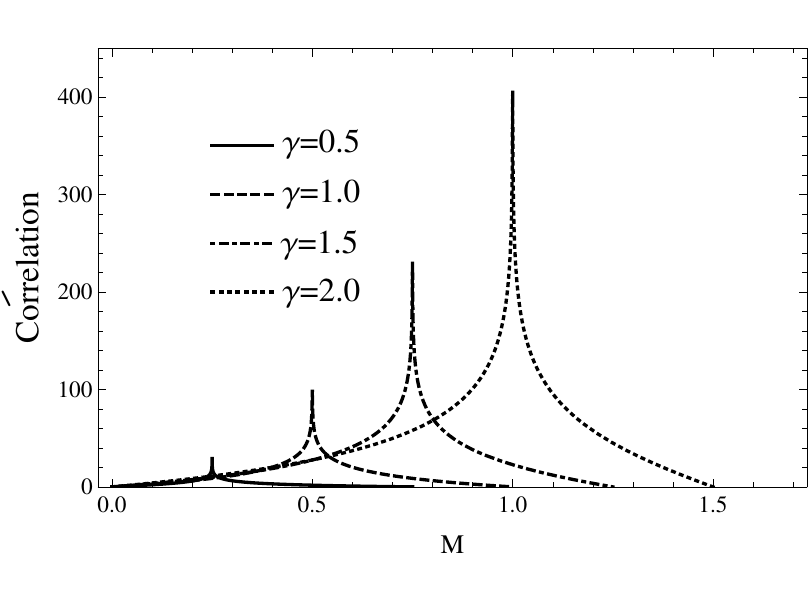}
\vspace{-0.2cm}
\caption{The correlations of particle emission.}
\label{correlacao}
\end{figure}

Finally, we analyze the graphical behavior of the correlation function of the emitted particles shown in fig. [\ref{correlacao}]. If the emitted particles have the same energy, the correlation will always be well located around $M=\gamma/2$. This effect is not observed for particles with different energies since, in this case, the correlation diverges for $M \rightarrow \gamma/2$. We also observe that the correlation increases as we increase the value of the rainbow parameter. This indicates that the rainbow parameter plays an important role on the correlation between emitted particles.

\newpage
\section{Concluding remarks}


Throughout all the paper, we can have an idea of how meaningful the rainbow parameter is for the thermodynamic properties. It not only modifies these quantities, but also contributes with new consequences, specially when it comes from the correlation between emitted particles. As observed in the last section, these consequences are particularly significant in the limit $M \rightarrow \gamma/2$, for which we have an increasing entropy variation (Figs. [\ref{variacaoent1}] and [\ref{variacaoent2}]) and, consequently, an increasing correlation (Fig. [\ref{correlacao}]). However, the contribution of the rainbow parameter is negligible for $M>>1$, which indicates that the rainbow gravity effects are more perceptible near the last stages of evaporation of the black hole.

Notably, the correlation has its maximal value in $M = \gamma/2$, suggesting that the particles emitted from the black hole are highly correlated moments before the black hole stops radiating. This is a striking result because, since the Rainbow Gravity restrains the complete evaporation of the black hole, we would expect the emitted particles to be uncorrelated, allowing the black hole information to be stored in its remnant. We may conclude from this that the Rainbow Gravity allows some of the black hole information to flow during the evaporation process, while maintaining the rest of it stored in its remnant. How this information is carried away from the black hole or whether these results are valid for other Rainbow models would require more investigation.

Finally, analyzing the entropy behavior of the model, we note that for $M>>1$, the entropy correction by the back-reaction effect is negligible, so $\Delta S\rightarrow 0$ when $M\rightarrow +\infty $. We emphasize that the back-reaction effect is considerable only for the $\frac{\gamma}{2}<M<\gamma$ region. This way, we conclude that the rainbow parameter must be taking into account only for values of $M$ for which the back-reaction effects are valid.

\section*{Acknowledgments} 
The authors thank the Coordena\c{c}\~{a}o de Aperfei\c{c}oamento de Pessoal de N\' ivel Superior (CAPES), and the Conselho Nacional de Desenvolvimento Cient\' ifico e Tecnol\' ogico (CNPq) for financial support. CASA thanks CNPQ for his grant No. 308638/2015-8. The authors also thank the anonymous referees for their valuable comments and suggestions.


\begin{thebibliography}{99}

  
\bibitem{akiyama}
K. Akiyama, A. Alberdi, {\it et. al.}, Astrophys. J. Lett. {\bf 875} (2019) L4.

\bibitem{2}
J. M. Bardeen, B. Carter and S. W. Hawking, Comm. Math. Phys. {\bf 31} (1973) 161.

\bibitem{Chen}
Y.-X. Chen and K.-N. Shao, Phys. Lett. {\bf B} {\bf 678}  (2009) 131.

\bibitem{medved}
A. J. M. Medved and E. C. Vagenas, Mod. Phys. Lett. {\bf A} {\bf 20} (2005) 1723.

\bibitem{vitor}
V. Santos, R. V. Maluf and C. A. S. Almeida, Phys. Rev. {\bf D} {\bf 93} (2006) 084047.

\bibitem{exp}
C. M. Will,
Liv. Rev. Relativ. \textbf{9} (2006) 3

\bibitem{datatables}
V. A. Kosteleck\'y, N. Russell,
Rev. Mod. Phys. \textbf{83} (2011) 11.

\bibitem{kostelecky}
V. A. Kosteleck\'y,
Phys. Rev. {\bf D} \textbf{69} (2004) 105009. 

\bibitem{bluhm}
R. Bluhm, 
Lect. Notes Phys. \textbf{702} (2006) 191.

\bibitem{hooft}
G.'t Hooft, Class. Quant. Grav.  \textbf{13} (1996) 1023.

\bibitem{amelino}
G. Amelino-Camelia, J. Ellis, N. E. Mavromatos, D. V. Nanopoulos and S. Sarkar, Nature, \textbf{393} (1998) 763.

\bibitem{iengo}
R. Iengo, J. G. Russo and M. Serone, J. High Ener. Phys. {\bf 11} (2009) 020.

\bibitem{dsr1}
J. Magueijo and L. Smolin, 
Phys. Rev. Lett. \textbf{88} (2002) 4.

\bibitem{dsr2}
G. Amelino-Camelia, Inter. J. of Mod. Phys. \textbf{11} (2002) 35.

\bibitem{dsr3}
J. Magueijo and L. Smolin, 
Phys. Rev. {\bf D} \textbf{67} (2003) 044017.

\bibitem{Magueijo}
J. Magueijo and L. Smolin,
Class. Quantum Grav. \textbf{21} (2004) 1725.


\bibitem{inflation}
J. D. Barrow and J. Magueijo, Phys. Rev. {\bf D} \textbf{88}  (2013) 103525.

\bibitem{inflation2}
G. Amelino-Camelia, M. Arzano, G. Gubitosi and J. Magueijo, Phys.Rev. {\bf D} \textbf{88} (2013) 041303.

\bibitem{branes}
A. Ashour, M. Faizal, A. F. Ali and F. Hammad, Euro. Phys. J. {\bf C} \textbf{76} (2016) 264.

\bibitem{wormholes}
R. Garattini and F. S. N. Lobo, \textit{Gravity's Rainbow and traversable wormholes},        
The Fourteenth Marcel Grossmann Meeting, (2017) 1448.

\bibitem{bb}
A. Awad, A. F. Ali and B. Majumder, 
J. of Cosmol. Astropart. Phys. \textbf{10} (2013) 52.

\bibitem{bb2}
G. Santos, G. Gubitosi and G. Amelino-Camelia,
J. of Cosmol. Astropart. Phys. \textbf{2015} (2015) 005.

\bibitem{bb3}
S. H. Hendi, M. Momennia, B. E. Panah and M. Faizal,
The Astrophys. J. \textbf{827} (2016) 153.

\bibitem{BHatLHC}
A. F. Ali, M. Faizal and M. M. Khalil,
Phys. Lett. {\bf B} \textbf{743} (2015) 295.

\bibitem{bhrg}
A. F. Ali, M. Faizal and B. Majumder, Euro. Phys. Lett. {\bf 109} (2015) 20001.

\bibitem{bhrg1}
Y. Gim and W. Kim,
J. of Cosmol. Astropart. Phys.  \textbf{2015} (2015) 002.

\bibitem{bhrg2}
S. H. Hendi and M. Faizal, 
Phys. Rev. {\bf D} \textbf{92} (2015) 044027.

\bibitem{bhrg3}
S.H. Hendi, M. Faizal, B. E. Panah and S. Panahiyan,
The Euro. Phys. J. {\bf C} \textbf{76} (2016) 296.

\bibitem{bhrg4}
S. H. Hendi, S. Panahiyan, B. E. Panah and M. Momennia,
Euro. Phys. J. {\bf C} \textbf{76} (2016) 150.

\bibitem{bhrg5}
Y. Kim, S.K. Kim and Y. Park, Euro. Phys. J. {\bf C} \textbf{76} (2016) 557.

\bibitem{bhrg6}
S. H. Hendi, B. E. Panah and S. Panahiyan,
Phys. Lett. {\bf B} \textbf{769} (2017) 191.

\bibitem{rainbowthermo}
Z.- W. Feng and S.-Z. Yang, 
Phys. Lett. {\bf B} \textbf{772} (2017) 737.

\bibitem{Remnants}
A. F. Ali, M. Faizal and M. M. Khalil, J. High Ener. Phys. \textbf{2014} (2014) 159.

\bibitem{Remnants2}
A. F. Ali, 
Phys. Rev. {\bf D} \textbf{89} (2014) 104040.

\bibitem{Remnants3}
A. F. Ali, M. Faizal and  M. M. Khalil, 
Nuc. Phys. {\bf B} \textbf{894} (2015) 341.

\bibitem{Hawking}
S. W. Hawking, Phys. Rev. D {\bf 13} (1976) 191.

\bibitem{Hawking1}
J. B. Hartle and S. W. Hawking, Phys. Rev. D {\bf 13} (1976) 2188.

\bibitem{Unruh}
R. Sch\"utzhold and W. G. Unruh, Phys. Rev. D {\bf 81} (2010) 124033.

\bibitem{Anabalon}
A. Anabalón, M. Appels, R. Gregory, D. Kubizňák, R. B. Mann, and A. Övgün, Phys. Rev. D {\bf 98}, (2018) 104038.

\bibitem{Sakalli}
I. Sakalli and A. Övgün, Europhysics Letters {\bf 118}, (2017) 60006.

\bibitem{Sakalli1}
I. Sakalli and A. Övgün, General Relativity and Gravitation, {\bf 48} (2016), 1.

\bibitem{Sakalli2}
I. Sakalli and A. Övgün,Eur. Phys. J. Plus {\bf 130}, (2015) 110. 

\bibitem{Ovgun}
A. Övgün and K. Jusufi, Eur. Phys. J. Plus {\bf 131}, (2016) 177. 

\bibitem{Sakalli3}
I. Sakalli, A. Övgün and K. Jusufi, Astrophys. Space and Science {\bf 361}, (2016) 330.

\bibitem{smolin} J. Magueijo and L. Smolin, Class. Quantum Grav. \textbf{21} (2004) 1725.
\bibitem{rovelli} C. Rovelli, Living Rev. Rel. {\bf 1} (1998) 1.
\bibitem{carlip}S. Carlip, Rept. Prog. Phys. {\bf 64} (2001) 885.
\bibitem{for} S. Forste, Fortsch.Phys. {\bf 50}, (2002) 221.
\bibitem{connes} Alain Connes, {\it Noncommutative geometry}. Academic Press, Inc., San Diego, CA, (1994).



\bibitem{Rovelli}
C. Rovelli, Living Rev. Rel. {\bf 1} (1998) 1.

\bibitem{Carlip}
S. Carlip, Rept. Prog. Phys. {\bf 64} (2001) 885.

\bibitem{Forste}
S. Forste, Fortsch.Phys. {\bf 50}, (2002) 221.

\bibitem{Connes}
Alain Connes, {\it Noncommutative geometry}. Academic Press, Inc., San Diego, CA, (1994).

\bibitem{a}
K. Srinivasan and  T. Padmanabhan, Phys. Rev. {\bf D} {\bf 60} (1999) 024007.

\bibitem{a1}
S. Shankaranarayanan, K. Srinivasan and T. Padmanabhan, Mod. Phys. Lett. {\bf A} {\bf 16} (2001) 571.

\bibitem{a2}
S. Shankaranarayanan, K. Srinivasan and T. Padmanabhan, Class. Quant. Grav. {\bf 19} (2002) 2671.

\bibitem{a3}
T. Padmanabhan, Mod. Phys. Lett. {\bf A} {\bf 19} (2004) 2637.

\bibitem{a4}
M. Angheben, M. Nadalini, L. Vanzo and S. Zerbini, J.  High Ener. Phys. {\bf 05} (2005) 014.

\bibitem{L2}
P. Li, M. He, J. C. Ding, X. R. Hu, and J. B. Deng, Advances in High Energy Physics {\bf 2018}, (2018) 1.

\bibitem{Refy}
O. El-Refy, S. Masood, L. G. Wang and A. F. Ali., Europhysics Letters {\bf 132}, (2020) 10006.

\bibitem{HUP}
R. J. Adler, P. Chen and D. I. Santiago, Gen. Relat. and Grav. {\bf 33} (2001) 2101.

\bibitem{HUP2}
M. Cavagli`a, S. Das, Class.  Quant. Grav. {\bf 19} (2004) 4511.

\bibitem{parikh}
M. K. Parikh, F. Wilczek, Phys. Rev. Lett. {\bf 85} (2000) 5042.

\bibitem{vagenas}
M. Arzano, A. J. M. Medved, E. C. Vagenas, J. High Ener. Phys. {\bf 09} (2005) 037.

\end{thebibliography}
\end{document}